# Prioritising Response-able IP Practices in Digitization of Electoral Processes in Africa

Angella Ndaka Ph.D.

Samwel Oando Ph.D.

Eucabeth Majiwa Ph.D.

**Abstract**

Globally, people widely regard technology as a solution to global social problems. In a democratic society, its citizens view technology as a way to ensure commitment and sustaining the nation's democracy by allowing them to participate actively in the democratic process. However, despite the hype surrounding technology and development, many developing countries still experience democratic challenges. The democratic challenges have further led to barriers that shape the political landscape, resulting in delusion, disappointment, and failures in the democratic and public good processes, such as the electoral process. This paper explores the relationship between intellectual property (IP) practices and the adoption of digital technologies used in democratic electoral processes. Specifically, it examines how the prioritisation of IP by technology service providers can disrupt socio-material relationships in democratic electoral processes and outcomes. Because of the hard boundaries associated with IP it creates an environment where the systems are controlled solely by technology IP owners, while the consequences of electoral processes are borne by citizens. This questions the response-ability and trust-ability of digital technologies in running democratic processes. Drawing from the parallels in Kenya's general elections of 2017 and 2022, this paper illustrates how IP practices form a hard boundary that impels technology owners to micromanage electoral processes, leading to tensions that potentially create conflict. This finding can be used by decision-makers to adopt digital technologies and protect IP without compromising electoral processes and disrupting relationships in the wider society.

**Introduction**

Many African countries have simultaneously experienced both democratic expansion and recession across the continent (Zeleza, 2019). These contradicting democratic directions sometimes occur in the confines of the same country as different segments of citizens engage in diverse collective efforts for what each state or group collectively considers as the 'most suitable' problem-solving 'democratic' initiative (Hendriks & Dzur, 2022; Zeleza, 2019). Some of the engagements include 'grassroots' mobilisation, publicising 'civic enterprises' by launching multi-million state projects, initiating 'self-help groups' for economic empowerment and rallying the ethnic blocs to 'vote for a share' in government, among other initiatives. However, in most cases, these engagements point to a static political mindset that connects democratic expansion to development based on false colonial legacies (Hendriks &

Dzur, 2022). Recently, the new forms of hyped civic engagements and enterprises have foregrounded the use of technology to achieve democratic ends. Further, technology has been positioned as an inevitable mythical solution to global social problems, including elections (Fauset 2008). However, by justifying superior technology strategies and policies that promote technological capacities, particularly the ability of technology to participate in decisive roles, these initiatives seem to promote capitalist and other political agendas (Cao, 2004; Cheng et al., 2019; Ibata-Arens, 2020). While they typically present themselves as missions of civilisation and modernisation, their concealed agenda is what some scholars call political and cultural dominance (Ibata-Arens, 2020; McDougall, 2017) disguised in the techno-fix agenda. Coincidentally, many developing countries continue to experience democratic challenges despite the subsequent hype on development and technology (Zeleza, 2019). These new forms of civic enterprises seem to have created barriers, which have subsequently contributed heavily to shaping the 'intellectual' political landscapes towards delusion, disappointment, and failures in some democratic and public good processes, such as the electoral processes. In most instances, these developments have emerged due to policies that provide short-term incentives for private companies while producing long-term socio-material consequences for vulnerable communities (Stevens, 1990). Despite concerns about these techno-social initiatives by affected communities, a push for a singular utopian future with shared visions - defined by a few entities – continues to thrive (Mukharji, 2012) and keeps pushing for unlimited technological growth at the expense of the user communities.

In a democratic process, the character of technology-related power relations and dynamics remains relevant and can thus contribute to the democratic quagmire and electoral failures. It is notable that electoral failures have caused disillusion among citizens, and elections have become a great determinant of existing 'peace and security' in every electoral cycle of many countries and in the regional network of states. In fact, the focus by many states has shifted to management of the electoral process as a priority to the development question (Otieno, 2018). It is in this systemic shift that in a democratic process technology has been positioned as an ultimate 'cure' for securing the citizen's commitment "to sustaining the nation's democracy" and to sustain the growing interest of citizens to participate actively in the democratic processes of their country (Zeleza, 2019, p. 159).

However, this happened when Africa's technological growth was still struggling and was at its infancy stage. This created an opportunity for others - especially the developed nations to adopt a nationalistic focus on technological development through marketing their tech innovations with little or no care about effectively addressing critical global issues like conflict, poverty, and inclusion (Duara, 2019). Some studies insinuate that technology is a political tool that is used to enact political goals for centuries (Latour, 2011; Hecht, 2019). Further, technology can support dominant power structures, further entrenching capitalist norms and corporate power in the global systems (Fauset, 2008). At the macro level, information about what technologies and innovations nations are coming up with remains so sifted – with most of it IP protected – in endeavours to protect national strength under the new techno-nationalistic policy regimes (Rasser & Lamberth, 2020). This creates forms of exclusivism and competitiveness, characterised by self-other relations, and is expressed differently by different countries (Duara 2019). This scenario has also created political barriers and other power structures that impact how global efforts and standard-setting for ethical and responsible innovations emerge (Latour 2011; Hasselbalch 2021). While some scholars call for a new policy regime for jointly formulating and implementing tech governance, these regimes are still vulnerable to subversion by hidden political, corporate and capitalist forces (Hochschild 2019; Barry 2021). One of the power structures created and maintained by the subsequent technological regimes is Intellectual Property practices – a subject we shall discuss at length in this chapter.

This chapter explores the relationship between intellectual property (IP) practices and the adoption of digital technologies to enhance the active participation of citizens in their respective democratic electoral processes. More specifically, it explores how prioritising IP rights by the technology service providers causes disruptions to socio-material relationships in democratic processes and electoral outcomes. We argue that the hard boundaries associated with IP create an environment and conditions through which the systems are, often, controlled solely by the technology IP owners (Carolan, 2017) while the consequences of electoral processes are borne by the citizens (Kutor, 2014). This puts into question the response-ability and trust-ability of digital technologies in running democratic processes.

The chapter is organised into five key sections. The first section commences by discussing Kenya's electoral cycles and how the process have been marked by a lack of accountability and challenges in the digital system. The second section engages with multiple barriers in intellectual property (IP) practices, specifically concerning the verifiability of election results. The third section highlights the different aspects of a compromised agency in digital technology for elections and the consequences, as seen in the 2017 and 2022 electoral processes, where IP-engineered power structures undermined the country's peace landscape. Last, the chapter envisions a repositioning of IP practices to be crucial, as technology should not be seen as a fixer for social problems without considering alternative worlds. In conclusion, the chapter argues for the need to address these issues as an essential step to ensuring a fair and transparent electoral process in Kenya and beyond.

**Elections and Technological Regimes in Kenya**

Kenya's electoral cycles, which adopted technology in three subsequent elections between 2013 and 2022, offer a good example of the possible tensions around the use of technology and the resulting contentions about the outcomes. For example,

> The unprecedented annulment of the August 2017 Presidential Election [was] the first time in the country's history. […] a presidential election was revoked by the judiciary, [*being*] the fourth time in the world, *that* marked a watershed moment that signalled to some the potential maturing [*democracy*] (Zeleza, 2019, p. 159; *emphasis added*).

The subsequent and related contentions since the 2013 elections demonstrate the direct impact of technological misuse or failures on the national peace and security architecture. The post-election turbulence and the associated violence rest on the citizen's hope that their collective 'standards of living will improve' if the outcomes favour their respective groups (Ahere, 2012; Otieno, 2018; Zeleza, 2019). However, any undetermined outcomes or any mistrust in the results of an electoral process arising from technology failures do not only lead to disillusionment but also come with anger and blame on the technology used. Zeleza (2019) argues that the disillusionment 'does not mean Africans do not want democracy', but rather, it is an indication that citizens want a 'democratic order' which delivers on their expectations. This can only happen when the citizens trust the technology used and when the state takes responsibility for the electoral outcomes resulting from the use of technology.

The development towards using technology to deliver a free and fair election in Kenya has led to the frequent call for mass voter re-registration in every election cycle using technology. The aim was to capture every voter's biometric identity details in 2013 in the new digital system in preparation for the general elections (Ahere, 2012). The introduction of digital registration brought great excitement, especially among the youth, and saw a record increase in the number of voters. AfriCOG recorded 14,340,036 voters registered at the end of December 18, 2012, *"representing approximately 80 per cent of the overall target population"* (AfriCOG, 2013, p. 5).

After the general elections of 2017, there emerged claims of digital citizen manipulation through what other commentators referred to as psychological warfare and an algorithmic interference of the system that led the ruling party and the supporting blocks to win with a visible algorithmic pattern[1]. This formed the beginning of political protest about '*fungua server'* (open the server), which reappeared again in 2022 when the company that was managing the digital parts of electoral processes could not verify who actually won the elections due to claims of hacking of the results transmission technology. At some point, after the 2022 general elections, the electoral commission faced technical challenges in explaining some obvious digital irregularities, which prompted the Supreme Court to request for servers to be opened. In response to the countrywide public demand, the company that provided electoral technology claimed that the server could not be opened because such an act would infringe the company's intellectual property (IP) rights (Smartmatic 2022). In a press release, the company retorted that,

> *"As per your request regarding the provision of image of NTC servers, we would like to clarify that such images contain software owned and copyrighted by Smartmatic and is thus protected. Providing full access would infringe our intellectual property rights"* (Smartmatic, 2022).

These contradictions lead us to ask the following pertinent questions: Who owns the electoral process and the data held in those servers? What level of control (agency) can citizens and end users, being the electoral commission, may be granted access to the data and the digital activities within the invisible infrastructure of a digital process? What are the possibilities and benefits of access to the server – i.e. should the servers be opened or not? How can electoral institutions navigate these tensions in the future to allow for ownership and active participation in the digitised electoral process by citizens? And, finally, what are the socio-material consequences of maintaining the digital power structures in the electoral process?

---

[1] https://www.theelephant.info/features/2019/08/09/cambridge-analytica-and-the-2017-elections-why-has-the-kenyan-media-remained-silent/

**The History of Electoral Malpractices and Violence in Kenya**

The Centre for Democracy and Development (2017) documents how democratic spaces have been affected by political and electoral violence over the past decade. This provides a reflection on the situation in Kenya during the past three consecutive election cycles of 2013, 2017 and 2022. Having faced challenges with the digital systems of elections, which came as a solution to the 'analogue' (manual) voting process, political players have resorted to frequent intimidation of political opponents and fears of constitutional amendments (Otieno, 2018). Political leaders in different camps, trying to perpetuate themselves in power or to access power, engage in polarised contestations which threaten to deny the citizens their rights to exercise sovereignty (Centre for Democracy and Development, 2017). These contestations use 'rigging of elections' as the claim for their polarised political positions, which often leave their supporters engaging in sporadic post-election violence (Yamano et al., 2010).

Political conflicts related to the flawed electoral processes were most common in Kenya with the introduction of multi-party politics since the year 1992 (Muriithi, 2010). Before then, the *mlolongo* elections (queuing system) of 1988 marked a long walk of rigging that commenced at independence in 1963 as an 'official currency' of elections malpractices under the one-party rule (Kamau, 2022). The rigging that happened during the period before the multiparty politics did not get much publicity, first due to the gagged communication systems and second due to the public fear of brutality meted out by dissenting voices by the Kanu regime. The citizens feared expressing discontentment in public. Hence, the claims of massive 'election irregularities and vote rigging, which polarised the country between 1992 and 1997, can be blamed for intensive ethnic tensions that followed with inter-ethnic violence (Endoh & Mbao, 2016; Kamau, 2022). Disgruntled parties, mostly organised by the opposition parties, took their protests to the streets, occasionally turning violent after security operations were deployed to counter the protesters (Otieno, 2018; Waweru & Ndirangu, 2017). It was not until the 2007 elections that Kenya suffered its worst humanitarian crisis since independence following the flawed presidential elections (Endoh & Mbao, 2016; Yamano et al., 2010).

> Upon the release of the election results, the opposition rejected the outcome on the grounds that the result was a product of fraudulent manipulations masterminded by the ruling party, including the cabinet, as they had access to the entire electoral process. Following this allegation, the opposition mounted a protest that devolved into widespread violence and resulted in the deaths of many people (Endoh & Mbao, 2016, p. 276).

Many actors in the governance debate have often assumed that different events of election-related violence result from a 'spontaneous reaction' to the election results (Endoh & Mbao, 2016). However, new evidence has continuously revealed how these political conflicts are connected to the promise of 'development', further linked to the fallacy of 'improved livelihoods' (Klopp & Kamungi, 2008; Rowan, 2022). Communities and political players have often confessed that violent conflicts after each election cycle stems from cumulative frustrations resulting from mismanagement of elections and finally leaving significant proportions of the national population with a feeling of being voiceless majority (Kimani, 2018; Klopp & Kamungi, 2008). The frustration from being voiceless (denied agency to be heard through the ballot) triggers anger and a sense of collective memory about 'historical disenfranchisement', which feeds into the perennial conflicts.

The realisation of the need to restore the voters voice at the ballot ushered in the calls for a 'digital' system as opposed to the 'analogue' system of election that was used from 2013 to 2022. For example, a shift to technology use to conduct elections was aimed at eliminating previous challenges, like double registration of voters and enhanced transparency in the transmission of results (Otieno, 2018). This intention is clarified by the observation that,

> Digital technologies for elections were introduced in Kenya with a vision that they would bring election reforms through increasing administrative efficiency, reducing long-term costs, and by enhancing transparency in the electoral process would enhance citizenry inclusivity (Omwoha, 2022, p. 147).

Oguk and Juma (2020) clarify that 'verification of electoral processes' was the most significant goal of shifting to the digital system of elections, as emphasised in the Constitution of Kenya 2010. As a priority, the elections laws enacted after the 2007-8 post-election violence required that "in every election, the given electoral commission should ensure that electoral systems used to be simple, accurate, verifiable, secure, accountable and transparent" (Oguk & Juma, 2020, p. 148). However, Omwoha (2022) presents a 'paradox' in the use of technology in the conduct of the past three elections.

The paradox is a manifestation of the fact that digitisation did not only subvert the democratic intentions for elections in Kenya but also failed the test of verifiability and transparency. According to Omwoha (2022), since 2013, the voting exercise has never taken place 'without a hitch'. For example, in 2017, the opposition party dismissed the general elections as a 'fraud', leading to the nullification of results by the Supreme Court. In general, the digital system of elections has not only failed but has disappointed the citizens as the results get repeatedly 'dismissed' by different stakeholders calling on the elections body "to open the servers" (Omwoha, 2022, p. 1). The opposition supporters renewed their protests, engaging 'in regular weekly demonstrations challenging the legitimacy" of the election outcomes (Allen et al., 2023, p. 5). This reveals a continued sense of failed promises by the electoral commission 'that counting, transmission and verification of electoral results would promote citizens' rights during the electoral process' (Omwoha, 2022; Otieno, 2018). The next section thus examines the dynamics of technology, which we can rely on to answer the questions about who owns the electoral process and the data held in those servers, given the situations witnessed in the past three elections.

**Technology and Antagonistic Power Asymmetries**

Whereas 'accountability of leaders to electorates remains a fundamental tenet of democracy', evidence from research shows that such accountability is not in the vicinity of Kenyan elections due to the technical challenges exhibited in the digital system (Otieno, 2018). This is worsened by the technical hitches in the 'verification' of the entire process to help build confidence in electoral outcomes (Oguk & Juma, 2020). The challenges experienced in the past can be attributed to the fact that science and technology are at best, hierarchical practices that privilege some assertions as objective realities over others, with a possibility of reducing large masses to objects of property and subjects of politics (Coombe, 1991). Therefore, there are many domains in the intersection of citizenship and civic and scientific agendas, where power struggles are real, and power relations are continuously and actively being challenged (Liboiron, 2017). For instance, in domains that are directly related to public goods like education, health, government service delivery systems, and democratic processes like elections, referendums, and online public participation systems.

This article, particularly, seeks to highlight the power dynamics within the conditions of access to data by citizens and other expert actors in civic technology. We question the extent to which citizens can be allowed to access their own data and how the level of access allowed by the systems management impacts the socio-material relationships in democratic electoral processes. Essentially, IP practices in technology-mediated electoral processes create barriers and a narrowed space for citizens to interrogate the accountability of the electoral processes critically. These technicalities affect the body and are mandated with the responsibility to manage the elections as they are often forced to hire 'external experts' – mostly foreign companies – who become the ultimate process owners and decision shapers of the process outcomes. We argue that whether or not citizens agitate for a transformation of power structures within civic tech, those structures rarely define particular practices, including citizen's agency, in a way that citizens' alternative views are enabled, creating an opportunity for credible and accountable civic techno-science (Liboiron, 2017, Wylie et al. 2014). Such opportunities come with power structures that include IP systems and practices. In such cases, discussions about power are not only explicitly discussed about civic technology but are made one of the main goals responding to the why and how technology is designed and adopted within democratic processes (Liboiron 2017).

Where technology is used exclusively to mediate public service delivery, that technology becomes an enabler or a barrier to effective and accountable service delivery to the people, due to existing power dynamics between different actors (Rosendahl et al. 2015; Ndaka & Majiwa 2024). It is possible, therefore, that the earliest digital interventions in the electoral process exhibit clear challenges in the inclusion of 'lay people' in the design of technology used. This issue is further characterised by antagonistic and asymmetrical power relations (Dad & Khan, 2023).

By interrogating how intellectual property has been applied within the intersection of civic technology, we question already how power manipulations are exercised in the management of democratic electoral processes. The struggle around IP and the deployment of civic technologies is not new (Wylie et al 2014; Scassa & Chung, 2015). The different and sometimes conflicting agendas create power asymmetries and struggles that sometimes promulgate legal, ethical and moral issues. Such asymmetries and struggles may relate to 'ownership' and 'custodianship' of knowledge, data, and information - which not only includes proper data collection practices, but also appropriate norms for data sharing and dissemination, and accountability of knowledge to those who contribute it (Scassa & Taylor, 2017). In cases where citizens provide data, respectful incorporation and sharing of this data may challenge conventional rules that regard to sharing that data in open sources. On the other hand, where conventional power structures like IP are in place, this may challenge the civic duty of citizens to critique data and related processes, further compromising how different parts of the society hold their democratic institutions accountable (Liboiron, 2017). Below, we explore in detail the intersections between intellectual property of technology and the democratic process of elections.

**Dynamics of IP, Civic Tech and Democratic Electoral Processes**

Intellectual property rights remain important in society due to their central role in promoting trade, investment and economic progress (Ahmed Lahsen & Piper, 2019). For instance, Saha & Bhattacharya (2011) take IP to mean any concepts, innovations, and artistic expressions that the public is ready to grant the status of property. Singh (2008), on the other hand, emphasises that an IP would be of any nature and thus may include any original work of the human mind, whether it be artistic, literary, technological, or scientific, and the person who does this is known as an inventor or creator. Such creations may take many forms, such as patents, copyrights, trademarks, industrial designs, geographic indications, trade secrets or undisclosed information (Bader, 2023; Rout, 2018). Usually, an inventor or creator is granted a legal right to keep their creations secret for a predetermined amount of time—as determined by their field of expertise, and thus, this legal right is known as IP rights (Singh, 2008). In this case, the goal of IP rights is to grant certain exclusive rights to the inventors or creators of that property, allowing them to profit commercially from their reputation or creative endeavours. The necessary laws and administrative procedures in place thus allow the protection of an IP.

IP rights laws and administrative processes first originated in Europe in the 14th Century, seeing the beginning of the patent-granting trend (Saha & Bhattacharya, 2011). Historically, England was technologically more advanced than other European nations in several areas, which attracted foreign artisans on favourable terms (Saha & Bhattacharya, 2011). However, it was in Italy where copyrights were initially recognised (Saha & Bhattacharya, 2011). Specifically, Venice is regarded as the birthplace of the IP system because it was where most legal thought took place, leading to the creation of the world's first laws and institutions, which eventually spread to other nations (Bainbridge, 2009). Outside of Europe, the Patent Act in India dates back more than 150 years (Lerner, 2002). The first law was the 1856 Act, which established a 14-year patent term modelled after the British patent system (Pujari et al., 2024; Singh, 2004).

Today, the intellectual output produced in a nation may be formally registered according to the national IP laws of that particular country; thus, each specific country governs the terms of the IP acquisition, upkeep, and enforcement (United Nations, 2022). In addition, there have been other attempts, such as the World Trade Organization (WTO) - Trade-Related Aspects of Intellectual Properties (TRIPs), to codify those obligations concerning the protection of copyrights, trademarks, industrial designs, geographical indications, patents, semiconductors and undisclosed information inside an international framework so that advancements can be recognised on a worldwide scale (Nemlioglu, 2019). Recently, indigenous peoples, local groups, and governments in biodiversity-rich countries have been seeking IP protection for traditional forms of creativity (Robinson et al., 2017)

As much as IP rights are important and have the potential overall positive effect on innovation and growth, there is evidence in the literature that the implementation of IP rights in the TRIPS agreement and other trade agreements has raised a number of pertinent ethical issues (Neves et al., 2021). First, IP rights are established as a socioeconomic instrument to provide innovator companies with a brief monopoly that allows them to charge prices for their inventions that are far higher than their marginal cost of production (Sonderholm, 2010). This monopoly leads to the exclusion of some potential customers from the market, which could be referred to as the "exclusion problem" or the "access problem" (Selgelid, 2008). When the monopoly granted by intellectual property rights creates a problem of life and death, like the lack of access to life-saving medications, the problem of exclusion and access becomes an ethical and moral one.

Second, strong IP rights may create availability problems from stifled competition. The availability problem stems from scenarios such as the price of the new product may directly correlate with the incentive to innovate, which is established by the innovation incentivisation mechanism of IPRs (Selgelid, 2008; Love & Hubbard, 2007). Thus, profits are only generated from sales under an IPR-driven system. This implies a greater motivation to devote resources to a product's research and development process resulting in the product fetching a higher price in the market (Arora, 2009). Under such circumstances, a number of innovators may invest in markets that will reap the maximum benefits, particularly where buyers can afford it, which may make the product unavailable in markets where the majority are poor and hence cannot afford it. Strong IP rights, according to Falvey et al. (2006), give foreign companies a patent advantage, which initially turns them into a monopoly and thus lessens competition. This could lead to a decrease in consumer welfare and production below the level society deems acceptable.

Thus, valuing IP rights based solely on economic factors ignores the cultural, social, and political ramifications of these rights and potential outcomes, such as their impact on society (Coombe & Turcotte, 2012). The economic, corporate, and political interests facilitate unequal power relations that keep subordinating the interests of the larger society particularly those of the entities that own technology (Ndaka et al., 2024; Hasselbalch, 2021). In the age of AI and data-driven economy, technological artefacts, as well as data and technological products of big data, are increasingly 'becoming' owned property. Hence, parts of them are claimed already, or will soon be claimed as intellectual properties. It is thus necessary to understand the dynamic of barriers in the IP practices that connect to citizens' limited agency in exercising democratic rights through digital electoral systems.

**Barriers in IP practices and Citizen's Agency - Boundaries and fences**

Following the difficulties associated with the IP rights and verifiability of election results, many countries in Africa and beyond, which have deployed electronic voting systems in the recent past, have dealt with 'numerous allegations and press reports' about irregularities in the electoral process (Ayawli et al., 2015). However, it has been incredibly difficult "to assess the credibility of these charges" (Ayawli et al., 2015, p. 2285). Some of the reasons for the difficulty lie in the making of critical assessments linked to the technical barriers in the IP practices of this technology. Consequently, it is apparent that both the complainants and prosecutors of fraud associated with technology use do not have adequate capacity to conduct a comprehensive post-election audit (Ayawli et al., 2015; Cheeseman et al., 2018).

Cheeseman et al. (2018) demonstrates a "remarkable increase" in the number of countries deploying digital technologies in their electoral systems, especially in Asia and Africa. More specifically, more than half of the countries of Africa "now involve digital equipment of some form" (Cheeseman et al., 2018, p. 1397). However, many stakeholders including experts in technology have successfully painted a 'troubling picture of election security' in the countries where digital systems have been implemented by producing elaborate reports about 'malfunctions' of technology use (Ayawli et al., 2015). These technologies involve biometric voter registration and identification systems and, more significantly, electronic results transmission systems. Despite these reports, governments and tech proponents continue to push for technology's inevitability and rightness in addressing social issues.

This shows how tech development is premised on techno-optimism and techno-fix discourses, which maximise information imperialism to promote a framing that entrenches dominant power structures (Klein, 2014, Mclennan, 2015). This kind of thinking foregrounds technology as solutions to social issues (Fauset, 2008) while invisibly driving exponential tech growth, capacities and accumulation of tech breakthroughs for those in power (Krier & Gillette, 1985). This optimistic approach has been instrumentalized by governments and corporates to influence national and global affairs (Wilson 2017) and has been used to prioritize large scale deployment of tech to address socio-economic issues, including the new problems created by the technology itself (Antal, 2018).

The main driver for the radical shift in electoral processes for many democracies is the hope that 'digital technology' can fix election malpractices influenced by politicians and election officials, which constitute existential messes in the national electoral processes. The tech optimists conceptually entangle rationality with superstitious optimism, which avoids a socially situated mediation of problems and uses it to sustain the tech industry (Wilson 2017, Mol, 2002). In this framing, technology is accorded a mythical inevitability and rightness position, which resists the socio-material situatedness of societal issues and solutions (Fauset 2008). While masquerading as immaculate and neutral (Hasselbalch 2021), technology is therefore introduced to "compensate for the weakness of the state and to deter malpractice" (Cheeseman et al., 2018, p. 1398). Technology thus becomes a 'band-aid' solution (Fazey & Fischer 2009), which fails to address the root cause of electoral conflict leaving behind deleterious impacts in economies where it is deployed. Therefore, this techno-fix ideology - instead of solving electoral issues – becomes anathema to democracy enthusiasts in developing countries.

Contrary to the public framings and expectations, evidence has proved that technologies use and rely on very complex procedures that are not only invisible in sophistication (Dieffenbach et al 2022), but are intrinsically liable to break down (Cheeseman et al., 2018;

Mathe, 2020; Mosero, 2022) and/or exhibit other socio-ethical and socio-legal issues. For instance, emerging technologies like AI have been shown to carry biases that emanate from design and training data (Diefenbach et al. 2022; Rosendahl et al. 2015; Boulamwaini & Gebru 2018). In many instances, therefore, digital technology has not only failed to achieve the promised outcomes and hence failed to meet situated public expectations (Fazey & Fischer 2009). Infact, it has more often increased widespread mistrust with claims of rampant manipulation, including manipulation of electorates using misinformation and disinformation. More importantly, this indicates the power asymmetries between different technology actors - that rarely get explicit acknowledgement nor address during tech conceptualisation, especially through thoughtful stakeholder engagement (Ndaka & Majiwa 2024; Farr 2018, Neef & Neubert 2020). These power asymmetries are responsible for the barriers in digital technology – which have not only emboldened complacency towards the pre-existing manual systems of elections (Dill 2005) but also foreground the interests of technology companies by optimistically promoting unlimited technology growth (Krier & Gillette, 1985). This technically robs the citizens of their power to provide oversight – an aspect that favours those with political interests and corporate power. In this regard, the citizen's agency is conspicuously missing, which compounds public frustrations and threatens to invoke more conflicts related to the conduct and outcomes of an election process. Dill (2005) argues that,

> The winners of an election are usually satisfied with the outcome [*whatever the challenges are*], but it is often more challenging to persuade the losers (and their supporters) that they lost. To that end, it is not sufficient that election results be accurate. The public must also know the results are accurate, which can only be achieved if conduct of the election is sufficiently transparent that candidates, the press, and the general public can satisfy themselves that no errors or cheating have occurred. (Dill, 2005, *Founder of the Verified Voting Foundation and Verified: emphasis added*).

This raises an intrinsic challenge in the digital system that must question how high the barriers of IP should be if an electoral system has to remain 'trustworthy' within situated socio-material dynamics. IP practices are, however, rhetorically framed questions around specific pieces of intellectual property and are dominantly defined by dominant knowledge groups who have historically dictated how scientific processes and related practices take shape (Ayris & Rose, 2023; Rosendahl et al., 2015). These circles argue that low barriers offer low protection, while high barriers offer protection against competitors (Bartow 2007). While these barriers are metaphorically effective practices against imaginary competitors, these actions also impede other actors, like citizens and scientists working outside the well-funded spaces, from accessing the scientific tools (Liboiron, 2017). In the absence of critical verification of electoral results by those entrusted with this process, the citizens' trust in both the process and the tools that facilitate the process becomes a myth.

Furthermore, the barriers technically deny the voters and their respective representatives' access to data they legitimately own and have a role to interrogate. This creates antagonistic relationships, especially in processes of elections where power relations are dominantly asymmetrical and are continuously contested (Dill, 2005; Kutor, 2014). In such cases, where citizens bear the agency of electoral outcome through sovereignty as enshrined by the national laws and constitutions, the IP barriers technically eliminate the citizen's right to exercise oversight (Kutor, 2014). By extension, the barriers exclude or take advantage to manipulate the scientists who advise the government. It is arguable that a pragmatic node of 'trustability' (Witt & Schnabel, 2020) of digital technology lies in the permissions to access and tinker with such scientific tools, which is fundamental to ensuring a transparent and accountable interpretation of the processes. Yet, in many cases, intellectual property laws outlaw the desired access to and manipulation of such digital tools to meet the needs and values of the populations whose lives and expectations depend on the election process (Ashton 2008; Dad & Khan, 2023; Wylie et al. 2014).

*Compromised agency in the digital technology for elections*

Having engaged with the level of barriers and the subsequent shortcomings in the use of digital technologies to carry out elections, it is inevitable to ponder the questions about the level of control by citizens (agency) in an electoral process. This implies an interrogation of the extent of involvement of citizens and the end users, the electoral commission, who have access to the data and the digital activities within the invisible infrastructure of a digital process without sanctions from the technology producers. Witt and Schnabel (2020) posit that the scope of agency, despite being a significant societal norm and a 'perceptual dimension' of the 'African governance' system, has seldom received adequate attention in academic literature. Nonetheless, scholars are unanimous on general recognition that digital interventions in elections are part of political interventions laced with corporate interests and, hence, are by nature political undertakings which deserve adequate agency by the citizens (Biswas & Deylami, 2019, Latour 2014, Hasselbalch).

Cheeseman et al. (2018) argue that the use of digital technology, being a new mode of gaining political legitimacy, must itself be able to boost the legitimacy of the process – and, by extension, the legitimacy of the government elected through this new technology. This can be practical only if the proponents of technology enhance the space of involving the public and the experts serving public interests in managing digital elections. Enhancing the space of participation by the public expands the level of agency of the public. According to Cheeseman et al. (2018, p. 1398), the "digital technologies can be used in many ways in elections, including the new communication technologies such as WhatsApp". However, the elections management bodies are always in a questionable rush to adopt "new ways of running elections" without considering the usability, including citizens' user skills. These scholars argue for an expanded agency exercised voters by noting that,

> The hope is that new technology will enhance the electoral environment in three main ways: by making the functioning of the electoral commission more robust and efficient, by reducing the scope for electoral manipulation, and by generating greater clarity and transparency regarding election outcomes (Cheeseman et al., 2018, p. 1398).

Against this backdrop, Mathe (2020) argues that digital technologies used for elections can be as simple as using "the social networking sites (SNSs)", which have the potential to "promote democracy, discussion and the participation of citizens" and, therefore, improve the voter's experience. However, the electoral commissions have in contrast, opted for expensive and complicated technologies that eliminate the citizens' rights to participate in the entire process from 'electronic voter registration, voter verification, and results transmission' (Cheeseman et al., 2018; Mathe, 2020).

While ignorance about technology by the electoral commissions may have widely informed such unpopular decisions, evidence shows that corruption nuanced in political manipulations is the main driver to making choices of complicated and untransparent digital systems and technology. Hence, the digitised electoral systems in developing democracies have often suffered not only from "electoral manipulations due to authoritarian leaders" (Mathe, 2020, p. 126), but also from the intentional political manipulations that bangle the digital technology systems used in elections (Omwoha, 2022). In such scenarios, citizens' agency is compromised because it is harder to separate intentional political manipulation from genuine concerns about IP rights. In Kenya, it is common to find those who are disenfranchised chanting slogans like, 'open the server' popularly known as the '*fungua sava*' in protests, informed to suspicion that the incumbents (in Africa) always have hidden intentions, and can use anything(including technology protocols like IP) to circumvent the process of electoral accountability and, subsequently, violate the will of the people (Mathe, 2020; Omwoha, 2022).

Towards this end, there is a possibility of many electoral malpractices disguised behind IP practices in civic tech, which not only compromise the agency of citizens in verifying the legitimacy of electoral processes but also entrench existing power structures. Thus, those in power use technology as a means to achieve their political ends. In contrast, those in technology continue being corporate beneficiaries of these corrupt and powerful regimes. This compromised state mostly results in citizenry who do not trust their governments and any machinery that is procured by the government. The result is socio-material consequences like post-election violence and continued protection – which disrupt many developing nations' social and economic environments.

*Restricted agency*

Restricted agencies emerge where high IP barriers are developed with restrictions that are meant for data and IP protection in digital technologies. In this case, tech developers and owners have discretionary authority to make ethical decisions that determine some levels of access on behalf of the users (Griffin et al., 2023; Carolan, 2017). This has an extended impact that restricts the accessibility to the technology servers that hold critical answers to electoral conflict, hence creating a narrowed space for citizens to interrogate electoral processes. Since the voter registers contain the personal data of millions of citizens, it is necessary that the 'privacy of this data' is protected (Wolf, 2017). However, this leaves key ethical decisions in the hands of a few powerful people, with dire consequences to the socio-material relationships within electoral processes (Griffin et al. 2023, Ndaka 2023; Burch & Legun 2021). These restrictions were created "both legally and technically." (Wolf, 2017, p. 28). To protect IP, biometric data in registration and verification kits are mostly used in elections throughout the country, and access permission is only given to a select few involved in the management of digital electoral systems. In this case, the electoral commission may exercise agency on behalf of the citizens, but only narrowly because there are parts protected by IP rights that can only be accessed by authorised technology owners and tech company personnel (Carolan 2018).

To bring this to perspective, this restricted agency happened during Kenya's general elections in 2022. When there was public demand for the servers holding electoral results to be opened, the company that was providing the digital electoral systems used IP rights as a basis for creating restricted access to the data that would have resolved the then-electoral stalemate, and the continued mistrust on the government systems (Hurley et al., 2022). The following quote summarises the reasons for the decision made by the company owners to deny access to electoral digital servers.

> As per your request regarding the provision of image of NTC server(s), we would like to clarify that such images contain software owned, copyrighted and IP protected by Smartmatic. Providing full access would infringe our intellectual property rights (Letter from Smartmatic to IEBC, dated 31 August 2022)

In cases where the electoral body and owners (citizens) of the electoral process are forced to request permission from a third-party organisation (tech company) to verify the legitimacy of their processes, then there is a restricted and misplaced agency. Even in this case, this access was not granted, despite claims by the company that their digital systems improved audibility, claiming 2022 as the most transparent election in the history of Kenya[2]. This poses the question of who defines electoral transparency and for whom. This is because, despite this bloated claim, there were country-wide chants of mistrust in response to what transpired in Kenya's August 2022 general elections. This shows how misplaced scientific objectivity and trust in digital systems, including the power structures within those systems, directly impact the quality of socio-material relationships (Burch et al., 2022) between electoral actors and systems. This further affects the level of accountability in electoral processes. Furthermore, where privileged groups get to decide what is of value or not, who can get that value and to what extent, and impose those impressions on the larger populations (Carolan 2018), they dictate what parts of technology can be accessed, by whom and why – which not only fences out perspectives of other actors but also restricts their agency partly or entirely in critical civic processes like elections (Ndaka, 2023).

**Consequences: The case of electoral processes in 2017 and 2022**

This chapter argues that IP-engineered power structures create barriers that compromise and restrict the agency of citizens in electoral processes and thus have far-reaching impacts on the country's peace landscape. Engaging in digital technologies for elections with already predetermined ends leads to compromised agency, frustrations, and the fear of manipulation becomes widespread. Dill (2005) asserts, for example, that due to the technical and legal restrictions in the digital technologies for elections, "e-voting technology is extremely opaque" (Dill, 2005, p. 1). Unless the restrictions against 'the agency' of the public and the electoral management bodies are addressed so the public has an active role beyond service procurement processes, no one will be able to "scrutinise some of the most critical processes of the election" (Dill, 2005, p. 1). According to Dill (2005), there is no way of ascertaining accountability in the digital process for voter registration, ballot collecting, vote counting, and result tallying because all these processes are often "conducted invisibly in electronic circuits" with only a few authorised entities having access to the true character of the processes and outcomes.

---

[2] https://www.smartmatic.com/featured-case-studies/kenya-fostering-transparency-through-technology/

As observed by Soderberg (2010), IP laws always outlaw the desired access to digital tools to meet the needs and values of the voters. The IP rights therefore often peripheries the voters' voices. Hence, the public has no means to confirm whether the digital equipment used "have recorded their votes correctly, nor will they have any assurance that their votes won't be changed later" (Dill, 2005, pp. 2–3). In an examination of this trend of challenges, Zeleza (2019) expresses a concern that Africa might attempt to achieve democracy backwards. Kenya Law outlines some extensive problems from the 2013 general elections in Kenya, which is described in their report as 'The triumph of Murphy's law' (Andago, 2013):

> Laptops and cell phones used ran out of battery power; additionally, some polling stations (particularly in the rural areas) had no power outlets. Many poll workers forgot basic PIN numbers and passwords needed to operate equipment. Secure servers intended for results transmission were unable to handle the volume of data being uploaded, leading to the breakdown. An error with the results transmission system source code multiplied the actual number of invalid ballots by 8 (an '8x error') (Andago, 2013, p. 10).

These issues raised from the audit of the 2013 elections were not only damning but also left a bleak doubt whether the digital technologies can achieve the intended purpose of preventing manipulations of the electoral process or they seek to serve other new ends where these manipulations are enhanced in invisibility. This concern becomes a genuine source of worry, especially when digital electoral systems are introduced before adequate measures are put in place to secure political legitimacy. The outcome of lethargy in the preparatory stages cannot be without far-reaching consequences for the citizens. Subsequently, we interrogate the different aspects of the socio-material consequences of maintaining an 'opaque' system that perpetuates digital power structures laced with coloniality and knowledge imperialism in the electoral process.

The failures in digital technology for elections dim citizens' hope of achieving political rights by electing their leaders without manipulations. This comes with mixed fortunes of turbulent democracy, which can be seen in "the difficulties the Mo Ibrahim Foundation has had in awarding the Ibrahim Prize for Achievement in African Leadership" (Zeleza, 2019, p. 161). According to Zeleza,

> the prize that 'recognises and celebrates excellence in African leadership' has been awarded only five times since its inception in 2007 to date. Given that the prize could not be awarded, due to lack of a deserving recipient, in seven of the past twelve years is a clear demonstration that more citizens are sceptical of their elected representatives (Zeleza, 2019, p. 161).

Since elections are the critical foundation for the legitimacy of governments, a dwindling 'source of accountability to the people' emerges immediately after contested election outcomes (Menkhaus, 2017). This challenges the 'key metric of successes' in elections as a fundamental tool 'for addressing political pathologies' that cause violent conflicts and reprisals in many countries. Political scientists have thus blamed flawed election processes for the 'strain on fragile compacts among elites', which also stand as a 'potential trigger of ethnic mobilisation' that has resulted in armed conflicts and coups in many countries (Menkhaus, 2008, 2017). Some scholars argue that unless the challenges of the digital platforms for elections are streamlined to expand the agency of the voters in Africa, it remains difficult to tell whether "democracy is feasible in sub-Saharan Africa" (Bratton & Chang, 2006, p. 1055).

The failures in the 2017 and 2022 general elections fall short of 'classic social and economic preconditions for democracy', like the troubled shared sense of national identity. Rowan (2022, p. 32) observes, for instance, that election violence "is generally mobilised by elites who are motivated to achieve power and wealth". Consequently, the disputed Kenyan elections of 2017 and 2022 bring the fear of "democratic recession", which gives way to collective disillusionment of the citizens (Zeleza, 2019, p. 160). Hence, the prolonged fear of possible violent conflict in Kenya after every election cycle must be addressed by increasing the direct agency of the voters in digital technologies for elections. The interventions through the accountable process of elections have the potential to rally leaders in both the ruling party and the opposition to enhance their willingness to go through the steps leading to conflict termination (Kimani, 2018).

**Repositioning IP practices**

Technology has often been presented as a fixer for social problems and positioned optimistically to serve the interests of those in power. While doing this, there is a scanty consideration of the existence of alternative worlds where these solutions are imposed (Ndaka et al., 2024). Where these are considered, these worlds are problematically presented through demographic or knowledge data, mainly to suit perspectives of conventional knowledge systems (Jaldi, 2023; Jerven, 2023), while they continuously fence out rich perspectives with persuasive knowledge claims which have served these contexts. This kind of framing creates power asymmetry in the way solutions are framed, as well as in how they respond to social problems (Bardzell 2014, Poliseli & Leite, 2021; Felt 2017). Despite the hype, many technologies being applied in Africa have struggled to address electoral accountability. This is partly because Africans do not understand accountability emanating from invisibility but also because this invisibility creates a new layer of mistrust. Furthermore, who ends up defining accountability and transparency? Do these definitions mean the same in different contexts? Given the African history of electoral malpractices, it is fruitless to impose technology simplistically as a solution to the complex intersecting election issues.

We therefore propose repositioning IP practices to enable response-able technology to thrive. In the normative conceptualisation of responsibility, endeavours are put in place to minimise negative outcomes by considering both ex-ante and ex-post responsibilities (Rijswijk et al. 2021). However, this remains a static social designation which does not consider "relational awareness of who is responding to who/what" (Burch & Legun, 2021, p. 22). Burch and Legun (2021) further argue that response-ability, on the other hand, treats people as more than human actors (technology, IP practices and agreements) and entities in a complex relational entanglement. Response-ability is described by Higgin (2021) as the ability of knowledge or solutions to respond to worlds beyond one-self. By design, the technology carries assumptions and social inclinations of the designer and owners (Rosendahl et al. 2015) and therefore is able to enact exclusions based on those assumptions and embedded biases. This is because knowledge is an entanglement of multiple entities and agencies and, therefore, is specific for situated places, shaped by norms and practices that sustain those norms (Mol 2002). More importantly, knowledge heavily carries the interests of those who intend to persuade. Electoral technologies produced externally have been shaped by multiple actions, agencies, politics, and interests which do not necessarily represent the interests of the end users. IP is simply "fencing off of ideas" (Boyle, 2002 p.23) and it fundamentally creates boundaries (between the creators and users) and intellectual resources that are positioned to reflect Western perspectives of property rights. Therefore, prioritising IP as central in cases where electoral transparency is in question produces negative social consequences to communities served by these technologies because these practices often exclude other stakeholders of these processes (Marot et al. 2005). This shows how current IP regimes over-focus on the needs of tech owners at the expense of the needs of the wider society (Oxford Analytic 2021). Hence there is less likelihood of considering property 'rights' from the perspective of users.

Kenyans consider votes as their property and data that determine a lot; therefore, external entities cannot hold it in some invisible servers. Focusing on the everyday socio-material relationships of the Kenyan electoral landscape and actors can support technology owners and those who implement them in those situated spaces to shape their abilities to respond within those electoral environments, especially where multiple agencies are compromised (Liboiron 2017). This makes responsibility a dynamic everyday practice, where the tech owners' response-abilities can dynamically shift to respond to situated socio-material contexts, including to human actors involved (Burch & Legun 2021). In spaces where elections have historically been compromised, tech owners should be willing to reposition their IP practices to respond to the challenges of technology mistrust. Adopting rigid IP practices only creates rigid barriers that compound the existing culture of electoral mistrust, with potentially dire consequences like perennial post-election and ethnically engineered conflict in these contexts.

**Conclusion**

Globally, technology has been positioned as an inevitable mythical solution to global social problems. In the democratic space, technology has been positioned as an ultimate hope for securing the citizens' commitment "to sustaining the nation's democracy". Technology particularly helps to sustain the growing interest of citizens to participate actively in the democratic processes of their country. However, many developing countries continue to experience democratic challenges despite the subsequent hype of development and technology. This has created barriers that have contributed heavily to shaping the 'intellectual' political landscapes towards delusion, disappointment, and failures in some democratic and public good processes, such as the electoral processes. This chapter explored the relationship between intellectual property (IP) practices and the adoption of digital technologies used in democratic electoral processes. More specifically, it explored how prioritising IP by the technology service providers can cause disruptions to socio-material relationships in democratic electoral processes and outcomes using Kenya as a case study. Kenya's electoral cycles, which adopted technology in three subsequent elections between 2013 and 2022, offered a good example of the possible tensions around the use of technology and the resulting contentions about the outcomes. In order to promote citizen inclusivity, election reforms were first intended to be brought about by improving administrative efficiency, lowering long-term expenses, and promoting transparency through digital technologies during Kenyan elections. One noteworthy development was the demand for widespread, technologically assisted voter re-registration during each election cycle. However, there have been difficulties, disputes, and political confrontations along Kenya's path to using technology for free and fair elections. For example, during the 2017 general elections, claims of algorithmic meddling and digital citizen manipulation emerged. Concerns around accountability and information access were raised when the electoral technology business refused to expose the server due to intellectual property (IP) rights in response to public demands for openness.

Thus, although being motivated by the desire for reforms and transparency, Kenya's adoption of digital technologies for elections has run into a number of controversies. The efficacy of these technologies in promoting inclusive citizenship and free and fair elections has come under scrutiny due to technical difficulties, political manipulations and other power structures created by extant intellectual property practices. This intricate terrain highlights the necessity for a meticulous reassessment of and repositioning of the function and application of technology IP practices in political and democratic processes. To promote responsible technology in Kenyan elections, this paper suggests realigning intellectual property (IP) laws to the situated socio-material dynamics of electoral processes in Kenya. In order to help technology owners and implementers adjust in compromised electoral situations, ordinary socio-material links are highlighted, with an emphasis on both ex-ante and ex-post obligations. The recommendation is to reposition IP practices to address issues with technology mistrust by acknowledging response-ability as a dynamic practice responsive to particular circumstances and human actors. The argument is for flexibility in reaction to past compromises in election processes against strict IP rules that could exacerbate already-existing electoral mistrust.